\documentstyle[12pt,twoside,cmp209,epsfig,amsfonts]{article}

\def\vac#1{{\bf #1}}

\newcommand{\be}{\begin{equation}}
\newcommand{\ee}{\end{equation}}

\def\vac#1{{\bf #1}}
\def\bsigma{\mbox{\boldmath $\sigma$}}

\def\subsubsec#1{$\triangleright$ {\em #1.}}

\def\ket#1{\left| #1 \right\rangle}
\def\bra#1{\left\langle #1\right|}
\def\braket#1#2{\left\langle\vphantom{#1#2} #1\right.
\left| \vphantom{#1#2}#2\right\rangle}
\def\trace{\mathop{\rm Tr}\nolimits}

\DeclareSymbolFontAlphabet{\mathbb}{AMSb}
\newcommand{\field}[1]{\mathbb{#1}}

\title[Historical and interpretative aspects of quantum mechanics]
	{Historical and interpretative aspects of quantum mechanics: a
	physicists' naive approach\thanks{Talk given at the
{\em Mochima theoretical physics spring school}, June 20th-24th 2005, 
by C. Dufour.}}
\author[Groupe M]
	{B. Berche, C. Chatelain, C. Dufour, 
	T. Gourieux and D. Karevski}
\address{
	{Groupe M, }
        {Laboratoire de Physique des Mat\'eriaux, }
        {UMR CNRS 7556,}\\
        {Universit\'e Henri Poincar\'e,  Nancy 1,}\\
        {F-54506 Vand\oe uvre les Nancy Cedex, France}\\
	}

\begin{document}

\pagestyle{myheadings}
\markright{version of \today}

\maketitle

\begin{abstract}
Many theoretical predictions derived from quantum 
mechanics have been confirmed experimentally during the last 80 years. 
However, interpretative aspects have long been subject to debate. 
Among them, the question of the existence of
hidden variables is still open.
We review these questions, paying special attention to historical
aspects, and argue that one may definitively exclude
local realism on the basis of present experimental outcomes.
Other interpretations of Quantum Mechanics are nevertheless not excluded.

\ {History of science}, 
{01.70.+w}\ {Philosophy of science},
{03.65.Ta}\ {Foundations of quantum mechanics; measurement theory}
\end{abstract}

\tableofcontents

\section{Motivations}
According to Mermin,  
contemporary physicists come in two varieties: 
those who are bothered by EPR arguments and 
Bell's theorem and those, perhaps the majority, who are not. 
In general, physicists who are not troubled 
by the kind of arguments used in the orthodox 
interpretation focus their attention on the 
technical part of quantum mechanics and on its 
unquestionable successes.
This positive attitude leads most of the time to a 
kind of ``operational realism" and to a complete 
disinterest regarding other interpretations. 
The present paper is a collection of 
lectures given at the Mochima summer school 
and at the university Henri Poincar\'e in Nancy. 
The aim of these lectures was to engage students 
and colleagues full of certitudes, in a discussion 
on some of the arguments bothering some contemporary 
physicists.  It is a short, superficial and non-exhaustive 
review of the kind of questions that 
part of the community is dealing with. 

In orthodox quantum mechanics a system is fully characterised
by its wave function. However, in 1935 
Einstein, Podolsky and Rosen (EPR) argued that such a theory is not complete, 
in the sense that not every element of reality has a counterpart in the 
theory. 
Bohr's reply underlined the fact that the EPR argumentation is 
misleading, since the interaction of the system with the experimental 
set-up is not taken into account. 
We discuss firstly the reality assumption  in the celebrated
debate between Einstein and Bohr.

If not convinced by Bohr's arguments, one may invoke
additional parameters -- 
the so-called hidden variables -- in order to fully account 
for a system's properties.
Historically, short after the EPR paper, the quest for hidden-variable 
theories weakened because the theorems of von Neumann, Gleason and Kochen as well as 
Specker claimed the impossibility of constructing 
hidden-variable theories reproducing all the results of quantum mechanics. In 
fact, these theorems have been shown to imply that such theories must 
exhibit {\em contextuality} (the dependence of a given measurement on the 
experimental set-up), rather than imply that hidden-variable 
interpretations are untenable. The subsequent step was Bell's 
assertion that any realistic local hidden-variable theory must satisfy 
inequalities that are violated by quantum mechanics.  
The theorem of von Neumann and the Bell inequalities are 
detailed in the third section.

The fourth section summarises the long series of experiments performed --
mainly in quantum optics -- since the beginning of the seventies and recently. 
Step-by-step improvements of the experimental expertise 
concerning the creation and manipulation of quantum states have led to 
the realization of several generations of experiments. These successive 
experimental tests, 
focusing essentially on the
effort made to circumvent the locality loophole,
favoured orthodox quantum mechanics and strongly 
indicates that local hidden-variable theories can be ruled out. 

However the hidden-variable theories in general have not been 
disqualified. For example, the non-local Bohmian mechanics still holds. 
This theory, which can be considered as a successful construction of 
hidden-variable theory, is described in the fifth section. It is 
empirically equivalent to orthodox quantum theory. 

We stress that we made no attempt at an exhaustive reference to the 
(vast) literature in the field. Interested readers may refer to recent reviews,
e.g. Ref.~\cite{Genovese2005}.

\section{1935: the reality assumptions in Quantum Mechanics}
In a seminal paper published in 1935~\cite{EPR1935}, Einstein,
Podolsky and Rosen (EPR) pointed out a tricky question about the
epistemological interpretation of quantum mechanics. Their objection
is supposed to prove the incompleteness of the theory with 
respect to physical reality.
According to quantum mechanics, all the physical
available information concerning a system is encoded in its 
wave function. If that was not the case, as argued by EPR in their paper, 
one would have to search for a more satisfying (complete) theory
which would encompass quantum mechanics.

\subsection{EPR assumptions}
According to EPR, in a complete theory,  
``{\em every element of the physical reality must have a counterpart
in the physical theory}''. This condition requires an experimental 
specification in order
to identify such an ``element of the physical reality''.
Thus, EPR propose a sufficient criterion  for their
following argumentation: ``{\em if, without in any way disturbing a 
system, we can predict with certainty (i.e. with a probability equal to unity)
the value of a physical quantity, then there exists an element of physical 
reality corresponding to this physical quantity}''.

Consider a particle in a momentum 
eigenstate, $|\vac p_0\rangle$ [i.e.
with a wave function $\psi\propto\exp(i\vac p_0\vac r/\hbar)$].
The momentum being $\vac p_0$ with certainty, it is 
then meaningful to consider
that it is an element of the physical reality 
associated with the particle.
On the other hand, the location of the particle 
is not determined in the
state $|\vac p_0 \rangle$, since all possible values 
of the coordinate are
equally likely. Certainly, one may proceed to a direct measurement. But
this measurement would perturb the state of the 
particle in such a way that
once the coordinate is known, the particle is no longer
in the state $|\vac p_0 \rangle$ and the precise 
knowledge of its momentum
is lost. One thus has to conclude that ``{\em when the momentum of a particle
is known, its coordinate has no physical reality}'', since it does not
satisfy the reality criterion. This example may be generalised to any 
observables $\hat P$ and $\hat Q$ which do not commute, since the 
corresponding physical quantities $P$ and $Q$  are 
subject to Heisenberg 
inequalities according to which one cannot know $P$ and $Q$ 
{\em simultaneously} with arbitrary accuracy.
According to EPR, 
two alternatives follow: either (1) ``{\em the quantum-mechanical
description of reality given by the wave function is not complete}''
or (2) [the quantum description is complete, but] ``{\em when the operators
corresponding to two physical quantities do not commute the two quantities
cannot have simultaneous reality}''.

\subsection{Incompleteness of quantum mechanics}
EPR consider that the second  alternative leads to an 
inconsistency. For this purpose,
assume a  system of two particles 1 and 2 initially
prepared without 
interaction, so that the wave function of the system is the product of two 
wave functions, the first one depending
on the physical properties associated with particle~1, the other 
depending on the physical
properties of particle~2. Suppose now that the two particles start to interact 
during a finite time. According to Schr\"odinger equation, the wave function 
of the system is known at any time during and after the interaction. But 
usually, due the interaction term, a mixing occurs and the wave function 
no longer appears as a product having the properties of the 
original one. The announced inconsistency arises from this entanglement of 
the particles state.
A simple illustration of such an entangled state has been given by David 
Bohm~\cite{Bohm1989},
using two-state physical quantities.
Consider a molecule
made of two spin $1/2$ atoms. The molecule is supposed 
to be prepared in a singlet state of  total spin, $s=0$, represented
by the entangled state 
\be|\psi\rangle=\frac 1{\sqrt 2}
(|\uparrow_z\rangle\otimes |\downarrow_z\rangle -
 |\downarrow_z\rangle\otimes |\uparrow_z\rangle),
\label{tg1}
\ee
where $|\uparrow_z\rangle\otimes |\downarrow_z\rangle$
is eigenstate of the observable $\hat S_{1z}$ with
eigenvalue $+\hbar /2$ and of observable $\hat S_{2z}$
with eigenvalue $-\hbar/2$~\footnote{We use the conventional notation
in which the left part of the tensor product refers to the first atom
while the right part to the second atom.}.
At a given time, the molecule breaks up
according to a process which does not change 
the initial total spin.

An experimentalist measures the spin
component $S_{1z}$ of atom 1. 
Suppose she gets the eigenvalue $+\hbar /2$.
According to the usual rules of quantum mechanics,
the state vector $|\phi\rangle$ immediately after the measurement is
the normalised projection of $|\psi\rangle$ on the eigenstate
$|\uparrow_z\rangle$ of  $\hat S_{1z}$,
\be|\phi\rangle=|\uparrow_z\rangle\otimes |\downarrow_z\rangle.\label{tg2}
\ee
As a result of the measurement on atom 1, the state vector is
disentangled and becomes a tensor product of individual state vectors
associated with each atom. By the way, the 
state being proportional to
$\dots\otimes|\downarrow_z\rangle$, the spin component 
$S_{2z}$ of  atom 2 takes the
value $-\hbar /2$ with probability 1.
If the experimentalist  had measured the 
spin component $S_{1x}$ of atom 1, and  had obtained 
the value $+\hbar /2$, then, immediately after measurement, the 
state vector would have become
the normalised projection of $|\psi\rangle$ on $|\uparrow_x\rangle
\otimes\dots$,  
\be|\phi\rangle
=\frac 1{\sqrt 2}|\uparrow_x\rangle\otimes\{|\downarrow_z\rangle-
|\uparrow_z\rangle\}
=
-|\uparrow_x\rangle\otimes|\downarrow_x\rangle.\label{tg3}
\ee
It follows that a further measurement of the 
spin component $S_{2x}$  of 
atom 2 leads to $-\hbar /2$ with certainty\footnote{One should have 
obtained this result directly, since the 
spin of the system 1+2
is zero. As a consequence, if the spin component of particle 1 
is $+\hbar/2$ along any direction,
that of particle 2 along the same axis is $-\hbar/2$.}.

We must conclude that 
in both cases, since no interaction has
perturbed atom~2 since the disintegration, the state 
vectors $\dots\otimes|\downarrow_z\rangle$
and $\dots\otimes|\downarrow_x\rangle$ must be associated with a 
unique element of reality
of this atom. As a consequence, 
according to the reality criterion
of EPR,  both spin components $S_{2z}$ and 
$S_{2x}$ should be simultaneously
associated with an element of reality concerning atom 2. 
Since it is well known that the observables 
$\hat S_{2z}$ and $\hat S_{2x}$
do not commute ($[\hat S_{2z},\hat S_{2x}]=i\hbar \hat S_{2y}$),
there is  an inconsistency and one 
should come back to the first
alternative: ``{\em the quantum-mechanical
description of reality given by the wave function is not complete}''.

In their original paper, EPR pointed out that one 
would not reach the 
same conclusions if, in the reality criterion, one had required
physical properties to be regarded as simultaneous elements of reality 
``{\em only when they can be simultaneously measured or predicted\/}''.
In this case, as shown in the previous example, the  
reality of
the physical properties depends
on the experiment achieved on atom 1, but as EPR concluded,
``{\em no reasonable definition of reality could be expected to permit
this}''.
This is nevertheless what Bohr would
accept.

In their original paper, EPR discuss the 
example of two
particles without interaction, prepared in a
coordinate-
and momentum- 
entangled state represented by the wave function
\be\psi(1,2)=\int_{-\infty}^{+\infty}\exp[i(x_1-x_2-X_1)p/\hbar]
\ \!d p, \label{tg4}\ee
where $X_1$ is a known constant. The contradiction 
lies in the fact that
a precise measurement of the momentum of particle 1, $p_1=p$, 
ensures that the momentum of particle 2 is known with certainty, $p_2=-p$,
while a precise measurement of the 
coordinate of particle 1, $x_1=x$, guarantees
that particle 2 has coordinate $x_2=x-X_1$ with certainty. 
Since particle
2 has not been perturbed since the preparation 
of the state, one concludes that
its momentum and coordinate are known with certainty at a given time, 
in contradiction with Heisenberg inequalities.

\subsection{Bohr's reply}
\begin{center}
\begin{figure}[!ht]
	\centerline{\hfill
	\psfig{figure=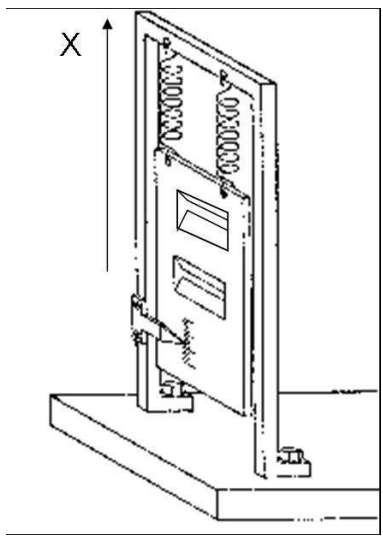,height=7cm}\hfill}
\caption{Double slit experimental setup.}
\label{Fig1}
\end{figure}
\end{center}

In his paper of 1935~\cite{Bohr1935}, Bohr 
reconsiders this example and 
elaborates upon the epistemology of 
{\em complementarity}, initiated during the
1927 Como and Bruxelles 
conference~\cite{Bohr1932}.
For that purpose, he points out that the EPR
example 
is realised with a two-slit Young device.
Consider two non-interacting particles 1 and 2 crossing
simultaneously a two-slit apparatus, not rigidly 
attached to the support as represented in Fig.~\ref{Fig1}.
One is then able 
to measure precisely the coordinate 
difference $X_1=x_1-x_2$ between the two 
particles at the time they cross it: this is the 
distance between the slits.
Measuring furthermore the spring elongation 
or compression due to the crossing
of the two particles, one is also able to 
deduce with arbitrary accuracy the
total momentum transfered to the aperture 
along the $x-$axis, $P_2=p_1+p_2$\footnote{Indeed, $X_1$ and $P_2$ 
are not conjugate and might 
be measured at the same time
with arbitrary accuracy. Considering 
pairs of conjugate variables 
$x_1,p_1$ and $x_2,p_2$ along $x-$axis, one may build other
 pairs of conjugate variables
$X_1=x_1-x_2$, $P_1=p_1-p_2$ and $X_2=x_1+x_2$, $P_2=p_1+p_2$.}
.
One has thus prepared the EPR entangled 
state  which corresponds to the particular case $P_2=0$.
A later precise measurement of the 
coordinate ($x_1$) or of the momentum 
($p_1$) of particle 1 enables one to deduce the coordinate ($x_2$) or
momentum ($p_2$) of particle 2 without in
any way perturbing it. But, and
this is where the reality criterion of 
EPR is unsuitable, one has to make 
a choice between the two experimental situations.
This choice requires an unambiguous use of 
the classical concepts needed for the
determination of the 
physical quantities. Measuring 
accurately $x_1$ for instance requires a well defined spatial frame
in the classical sense,  which needs the apparatus (a single slit
aperture for example)
to be rigidly 
fixed to the support that defines
that reference frame. But in doing so, one drops out 
further possibility to measure $p_1$, since 
the momentum transfer between the
particle and the aperture is lost in the 
support. How then can an element
of reality be associated with $p_1$, and thus with $p_2$, 
which are variables originating from
classical concepts while we are unable to  
perform any measurement of them
through the apparatus used, due to the very 
existence of the action
quantum $h$.
Measuring  $p_1$ 
on the other hand requires 
an experimental setup which enables to use classical conservation laws
of the momentum transfer between the particle 1 and the apparatus. This
may be for instance the same single 
slit aperture, but now dissociated from
the support, allowing for a measurement of the momentum through  
the elongation or compression of a spring.
In doing so, we loose the ability of an 
accurate measurement of the coordinate 
$x_1$ (and thus $x_2$), the precise measurement
of which requires a fixed frame
of reference.

Therefore, according to Bohr, the reality 
criterion proposed by EPR contains
an essential ambiguity when it is applied to 
atomic systems, ambiguity which
lies in the words ``{\em without in any way disturbing a 
system}''. Of course in the example of EPR, there 
is no mechanical perturbation
during the critical step of the measurement process, ``{\em but there is
essentially the question of an influence on the very conditions which define
the possible types of predictions regarding the future behaviour of the 
system.}''
According to Bohr, the EPR argumentation 
is misleading, since it is based on 
the assumption that one may intelligibly 
speak about the state of a physical
system, regardless of any experimental setup. 
This is equivalent to a 
discourse on the ``the thing in itself'', 
regardless of external conditions 
in which it is observed. 
According to Bohr, what quantum phenomena reveal, thanks
to Heisenberg inequalities, is precisely the 
inability to define properties 
(elements of reality) of a quantum system
without taking into account their interaction with other systems.
He  proposes a description of the quantum world
in terms of complementarity, a way which although discussed 
in terms of classical concepts, accounts for the 
fact that no separation between the behaviour of atomic systems
and their interaction with measurement devices 
is possible. 
This is complementarity. 
Wave and particle behaviours of light and matter are complementary
in the sense that when a measurement is performed, one should use both 
classical concepts when interpretating in a consistent way 
the observed phenomena as a whole.

\subsection{Bohm's further explanations}
Although it plays an important role in the 
argumentation of EPR, Bohr did not extensively discuss the wave 
function. We will follow Bohm 
to provide a complement to the orthodox interpretation
of that point and of the significance of 
observables of quantum mechanics.

Firstly, one observes that the EPR argumentation 
is based on two implicit assumptions which are never
discussed in the original paper. The first 
one is that the world may be correctly analysed
in terms of distinct elements of reality which 
have separate existence. This is the problem of
{\em separability}, to which we will return later. 
Secondly, elements of reality must have 
a definite mathematical correspondence in a complete
theory. This second statement is in fact only 
a stronger version of the definition of elements
of reality given by EPR, but it is 
this stronger version which is used in the paper, since 
the element of reality 
is defined even before the measurement.

If one rejects both assumptions, which are 
at the basis of any classical theory, the paradox
disappears. Indeed, orthodox quantum theory 
assumes that the strong correspondence stated above
emerges at the classical level only. At a 
quantum level, mathematical description is provided
through the wave function which is not in strong 
correspondence with the system, but in
a statistical correspondence, although the wave function provides the most complete possible
description of the system. In order to reconcile both seemingly
contradictory aspects of the wave function,
one has to accept that elements of 
reality of a quantum system do generally exist in an
unprecise way. These are potentialities, likely to be realised when 
the interaction of the system
with a classical measurement device occurs. 
The wave function
describes the propagation of these correlated potentialities.
Hence, concerning two non-commuting observables, 
such as the coordinate and the momentum of a 
particle, neither of them has a precise and 
definite existence. Rather, each of them is likely to 
be actualised when interactions with the 
adequate apparatus takes place. Within this approach, 
the characteristics attributed to the 
observables are not intrinsic to the 
subsystem, but rather through their interaction in the complete system.

\subsection{Einstein's point of view on separability}
In his correspondence and in further 
papers published after the EPR article~\cite{Einstein1989}, 
Einstein clarified his
point of view. The simple fact that the wave 
function describing particle 2 after measurement made on
particle 1 depends on this measure is, according to him, the sign 
of incompleteness of quantum theory~\cite{Einstein1948}:

{\it {}``From the point of view of the [orthodox] 
interpretation  this means that
according to the choice of a complete measurement on $S_1$ a 
different real situation with respect to $S_2$ is created, which
is described through differently natured 
$\psi_2$, $\psi_2'$, $\psi_2''$ etc.

{}From the point of view of quantum mechanics {\em alone} this does
not present any difficulty. According to the specific choice of
the measurement on $S_1$ a distinct real situation is created and
the necessity, to associate with the system
$S_2$ at the same time two or more distinct wave functions
$\psi_2, \psi_2',\ldots$, can never arise. 

However, the situation changes if one tries to maintain, 
simultaneously with the principles of quantum mechanics, 
also the Principle 
of the  independent existence of a real situation in two 
separated parts $R_1$ and $R_2$ of space. In our example a
complete measurement on $S_1$ means a physical interaction which
only concerns the part $R_1$ of space. Such an interaction, however,
cannot directly influence the physical reality in a part $R_2$ 
of space which is far away. This would imply that any assertion
with respect to $S_2$ which we could achieve through a complete
measurement  on $S_1$ must also be valid for the system $S_2$
when no measurement at all is made on $S_1$. That would mean that
all statement on $S_2$ must be valid which can be derived by 
specifying $\psi_2$ or $\psi_2'$ etc. This is of course impossible,
if $\psi_2$, $\psi_2'$ etc. stand for mutually distinct real facts
of $S_2$, that is one enters in conflict with the [orthodox] interpretation
 of the $\psi$-function. 

Without doubt, it appears to me
that the physicists which take the
description of quantum mechanics for definitive in principle, will
react to this thought as  follows: they will drop  the requirement
 of  the independent existence of the Physical  Reality present
in different parts of space; they are justified in pointing out
that Quantum Theory nowhere makes use of this requirement. 

I admit this, but point out: if I consider the physical 
phenomena I know about, in particular  those which are 
accounted  for so successfully  by quantum mechanics, I still find
nowhere a fact which would make it probable to me that the
requirement [of separability] should be abandoned. I am therefore inclined
to believe that in the sense of [orthodox interpretation] the  description of
quantum mechanics should be viewed as an incomplete and
indirect description of reality, which will be replaced  later
by a complete and direct one.'' 
}

\subsection{What orthodox physicists should accept, and 
what others may reject}
In 1935  Schr\"odinger  cat paradox also appeared. A cat enclosed
in a box plays the role of a measurement apparatus with two
macroscopic mutually exclusive  outcomes, dead or alive.
The answer is given in modern quantum 
mechanics through the concept of quantum decoherence of
entangled macroscopic states. This 
notion is a bit out of the scope of the present contribution~\footnote{Quantum 
decoherence is the phenomenon 
according to which the time evolution of a quantum 
superposition of states, when interacting with an 
environment, displays interference 
terms which are amazingly rapidly 
suppressed at a time scale which is by orders of magnitude
shorter than the typical relaxation times of the system}, 
but is required  
to justify epistemological completeness of orthodox
quantum theory. 

As a summary, an orthodox physicist has to 
admit that physical properties attributed to a quantum 
system are only potentialities which become physical 
reality when interacting with a
measurement apparatus. 
For such a physicist, the concept 
of trajectory of a particle 
is meaningless, unless a device is used to 
measure it. He (she)
also admits that ``nature plays dice'', 
that is  probabilities are inherent to quantum
phenomena.
Those who do not share this opinion can 
invoke some hidden variables at the origin 
of probabilities which would then reveal 
our ignorance of these variables. 
In this direction, one has to distinguish
between non-local hidden-variable theories 
(Bohm - de Broglie theory is the paradigmatic example)
and local hidden-variable theories which 
are discussed below in this paper. 
The question
of separability is nevertheless still present. 

\section{On the impossibility of local hidden variables}
\subsection{Hidden-variable theories}
In classical physics, measurement processes are intuitive: the result of
the measurement of any physical observable ${\cal A}$ is a function
of the state of the system, e.g. the positions $\vac r_i$ and the velocities
$\vac v_i$ for a set a particles. The measurement of ${\cal A}$ gives the
same result in all systems identically prepared. For this reason, the classical
state is said to be a dispersion-free state. In contradistinction, quantum
states are not dispersion-free. For systems identically prepared in a given
quantum state $\ket{\psi}$, the measurement of ${\cal A}$ does not give 
a unique result but one of the eigenvalues $a_i$ of the associated 
hermitian operator 
$\hat A$ with the probability
$|\braket\psi{\phi_i}|^2$ if $\hat A\ket{\phi_i}=a_i\ket{\phi_i}$.
When the number of such systems becomes large, the average measurement 
is thus $\langle A\rangle=\bra\psi\hat A
\ket\psi$ and the mean dispersion around this average 
is $\sigma_A=\sqrt{\bra
\psi\hat A^2\ket\psi-\bra\psi\hat A\ket\psi^2}$. The quantum state $\ket\psi$
is not sufficient to determine the output of a unique measurement of
${\cal A}$. According to the orthodox interpretation of quantum mechanics,
this uncertainty on the output of an unique measurement is a fundamental
property of quantum world.

It is tempting, as Einstein, Podolsky and Rosen did, to make
the assumption that the wave function $\ket\psi$ does not give a complete
description of the state of the system. There may exist a set of hidden
variables $\lambda$, not yet experimentally observed, such that 
$\{\ket\psi,\lambda\}$ 
completely determines  the output of a unique measurement, i.e.
that $\{\ket\psi,\lambda\}$ is a dispersion-free state. Let 
$V_{\{\ket\psi,\lambda\}}(\hat A)$ be the output of a unique measurement on a
system in the complete state $\{\ket\psi,\lambda\}$. The function $V$ maps
the dispersion-free states $\{\ket\psi,\lambda\}$ onto the eigenvalues of the
operator $\hat A$. Since the variables $\lambda$ are unknown,
one cannot prepare experimentally different systems in the same
state $\{\ket\psi,\lambda\}$. 
One can only ensure that the wave function
$\ket\psi$ is the same, but the hidden variables $\lambda$ remain 
uncontrolled for whatever reason~\footnote{They may depend on the history
of the system, non-separability with the rest of the Universe, 
intrinsic
stochasticity, etc. } 
in an ensemble of different systems.
Let $\wp(\lambda)$ be
the probability distribution of the hidden variables among the different
systems. When the number of such systems becomes large, the average
measurement of ${\cal A}$ should be equal to the prediction of the
quantum theory:
	\begin{equation}
	\langle A\rangle=\int V_{\{\ket\psi,\lambda\}}(\hat A)
	\wp(\lambda)d\lambda=\bra\psi\hat A\ket\psi .
	\label{VarC1}
	\end{equation}
Another important difference to quantum mechanics is that the
wave packet reduction postulate is not necessary anymore. In order
to explain the empirical fact that the same output $a_i$ is obtained when
repeating the same measurement ${\cal A}$ on the same system, one needs
to assume in quantum mechanics that the wave function $\ket\psi$ of the
system is projected onto the eigenvector $\ket{\phi_i}$ of $\hat A$
associated with the eigenvalue $a_i$. In hidden-variable theories, it is
sufficient to assume that the complete state $\{\ket\psi,\lambda\}$ is not
modified during the measurement process. That is the usual classical
measurement picture.

\subsection{von Neumann no-go theorems}
In the very early years of  quantum theory, von Neumann studied the
possibility of an underlying hidden-variable quantum 
theory~\cite{vonNeumann1932}. He showed that the linearity postulate of quantum
mechanics, implying
	\begin{equation}
	\langle \big(\alpha A+\beta B\big)\rangle
	=\alpha \langle A\rangle+\beta\langle B\rangle,
	\label{VarC2}
	\end{equation}
cannot be fulfilled by a hidden-variable theory, which demonstrates
according to him, the impossibility of such a theory. In the following, we give
an outline of his demonstration (for further details, see for instance
the excellent dissertation by D. Hemmick~\cite{Hemmick1996} on which this
section is based). First, von Neumann  showed that equation
(\ref{VarC2}) implies that $\langle A\rangle=\trace\hat A\hat\rho$ where
$\hat\rho$ is a density operator equal to $\hat\rho=\ket\psi\bra\psi$ in
quantum mechanics. Invoking equation (\ref{VarC1}), one can show that
(\ref{VarC2}) also applies to $V_{\{\ket\psi,\lambda\}}
(\hat A)$, i.e.
	\begin{equation}
	V_{\{\ket\psi,\lambda\}}(\alpha A+\beta B)
	=\alpha\ \!V_{\{\ket\psi,\lambda\}}(A)
	+\beta\ \!V_{\{\ket\psi,\lambda\}}(B),
	\label{VarC3}
	\end{equation}
and thus  $V_{\{\ket\psi,\lambda\}}(A)=\trace\hat\rho_\lambda\hat A$.
Applying this relation to the projector 
$\hat P_\phi=\ket\phi\bra\phi$, 
one obtains  
	\begin{equation}
	V_{\{\ket\psi,\lambda\}}(\hat P_\phi)=\trace\hat\rho_\lambda
	\ket\phi\bra\phi=\bra\phi\rho_\lambda\ket\phi.
	\label{VarC4}
	\end{equation}
The measurement output $V_{\{\ket\psi,\lambda\}}(\hat P_\phi)$ is simply
a matrix element of $\hat\rho_\lambda$. On the other hand, since 
$\hat P_\phi^2=\hat P_\phi$, 
and since repeating twice the measurement
should give twice the same output in a 
classical measurement process,
one has the relation
	\begin{equation}
	V_{\{\ket\psi,\lambda\}}(\hat P_\phi)
	=V_{\{\ket\psi,\lambda\}}(\hat P_\phi^2)
	=\big[V_{\{\ket\psi,\lambda\}}(\hat P_\phi)\big]^2,
	\label{VarC5}
	\end{equation}
which implies that $V_{\{\ket\psi,\lambda\}}(\hat P_\phi)$ is equal to $0$ or
$1$. The vector $\ket\phi$ being arbitrary, we are led to the conclusion that
all matrix elements of $\hat\rho_\lambda$ are either  $0$ or  $1$,
in contradiction with 
$\trace\rho_\lambda=1$.

The correctness of the von Neumann demonstration is not questionable but
one may wonder whether the postulate (\ref{VarC3}) is really necessary.
The alternative demonstration of the von Neumann theorem provided by
Schr\"odinger is helpful to address this question~\cite{Schroedinger1935}.
Consider a particle in a harmonic potential with Hamiltonian
        \begin{equation}
	\hat H={\hat p^2\over 2m}+{1\over 2}m\omega^2\hat x^2.
	\label{VarC6}
	\end{equation}
Equation (\ref{VarC3}) reads for $\hat H$,
        \begin{eqnarray}
	V_{\{\ket\psi,\lambda\}}(\hat H)
        &=&{1\over 2m}V_{\{\ket\psi,\lambda\}}(\hat p^2)
        +{1\over 2}m\omega^2V_{\{\ket\psi,\lambda\}}(\hat x^2)
	\nonumber\\
        &=&{1\over 2m}\big[V_{\{\ket\psi,\lambda\}}(\hat p)\big]^2
        +{1\over 2}m\omega^2\big[V_{\{\ket\psi,\lambda\}}(\hat x)\big]^2.
	\label{VarC7}
	\end{eqnarray}
Since in a given experiment $V_{\{\ket\psi,\lambda\}}(\hat x)$ and
$V_{\{\ket\psi,\lambda\}}(\hat p)$ can take any  
real value, it is not possible to ensure that
$V_{\{\ket\psi,\lambda\}}(\hat H)$ is always equal to one of the 
discrete eigenvalues
$\hbar\omega\big(n+{1\over 2}\big)$ of $\hat H$. Unlike von Neumann,
Schr\"odinger questioned the validity of the postulate (\ref{VarC3}).
Indeed, $\hat x$ and $\hat p$ are non-commuting operators, i.e.
$[\hat x,\hat p]=i\hbar$, which means, according to Heisenberg uncertainty
principle, that position and velocity cannot be measured during the
same experiment with arbitrary accuracy. 
As a consequence, $V_{\{\ket\psi,\lambda\}}(\hat x)$
and $V_{\{\ket\psi,\lambda\}}(\hat p)$ have
no simultaneous meaning.
Two different experiments have to be performed to measure both
position and velocity. The postulate (\ref{VarC3}) is not applicable
 to non-commuting operators but only to sets of
commuting ones, i.e. to quantities that can be measured during
the same experiment. Using the same approach than von Neumann, Gleason
showed that restricting the postulate (\ref{VarC3}) only to sets of commuting
operators leads to the same conclusion: the 
impossibility of such hidden-variable
theory~\cite{Gleason1957}. Kochen and Specker obtained the same
result in the case of
a spin one particle~\cite{KochenSpecker1967}. However, as latter recognised
by Bell, one can find operators such that $[\hat A,\hat B]=0$
and $[\hat A,\hat C]=0$ but $[\hat B,\hat C]\ne 0$ which means that both
${\cal A}$ and ${\cal B}$ can be measured with arbitrary accuracy with
a given experimental setup,
${\cal A}$ and ${\cal C}$ can be measured similarly
with a second experimental setup
but no experimental setup allows the simultaneous precise
measurement of ${\cal B}$
and ${\cal C}$. As a consequence, the mapping 
function $V_{\{\ket\psi,\lambda\}}
(\hat A)$ giving the output of a unique measurement of ${\cal A}$ should depend
on the experimental setup used. 
This obvious property
is called contextuality.

\subsection{Bell inequalities}
Consider once more Bohm's version of EPR experiment.   
The system is described in quantum mechanics by 
the singlet state (\ref{tg1}).
The spins are sent away from each other and are measured in two different
directions $\vac a$ and $\vac b$ at the same time, or at least at times
sufficiently close to forbid any communication (at the speed of light)
between the two measurements. The outputs $\sigma^{\vac a}_1
=\bsigma_1.\vac a$ and $\sigma^{\vac b}_2=\bsigma_b.\vac b$ of
the measurements are multiplied and averaged over a large number of
experiments, leading to a correlation function
$C(\vac a,\vac b)=\langle\sigma^{\vac a}_1
	\sigma^{\vac b}_2\rangle$.
Quantum mechanics predicts that this correlation function
is equal to
	\begin{equation}
	C_{QM}(\vac a,\vac b)=-\vac a.\vac b.
	\label{VarC9}
	\end{equation}
Note that if the two spins are measured in the same direction $\vac a=\vac b$,
the correlation is simply $C_{QM}(\vac a,\vac a)=-1$ which means that
in all experiments, the two spins have always been observed in opposite
directions, reflecting the fact that the total spin of the
system is zero. 

One is tempted to think in terms of classical vectors, but
Bell showed that this picture is
incompatible with the average correlation (\ref{VarC9}) predicted by
quantum mechanics~\cite{Bell1964}.
Indeed, in a {\em local} hidden-variable theory, the average correlation reads
        \begin{equation}
        C(\vac a,\vac b)=\int d\lambda\ \!\wp(\lambda)
        A(\lambda,\vac a)B(\lambda,\vac b),
        \end{equation}
where $A(\lambda,\vac a)=V_{\{\ket\psi,\lambda\}}(\sigma_1^{\vac a})=\pm 1$ and
$B(\lambda,\vac b)=V_{\{\ket\psi,\lambda\}}(\sigma_2^{\vac b})=\pm 1$. 
Note that
this expression satisfies the requirement of {\em locality}: the 
measurement
of $\sigma_1^{\vac a}$ for instance does not affect that of 
$\sigma_1^{\vac b}$~\footnote{In the case of simultaneous measurements,
a {\em non-local} hidden-theory correlation would
be written as 
$C(\vac a,\vac b)=\int d\lambda\ \!\wp(\lambda)
        A(\lambda,\vac a,\vac b)B(\lambda,\vac a,\vac b)
        $}.
Since the total spin is zero and  $C(\vac a,\vac a)=-1$, one can write
        \begin{equation}
        -1=\int d\lambda\ \!\wp(\lambda)
        A(\lambda,\vac a)B(\lambda,\vac a)
        \ \Leftrightarrow\ A(\lambda,\vac a)=-B^{-1}(\lambda,\vac a)
        =-B(\lambda,\vac a).
	\end{equation}
As a consequence, the difference of the average correlations measured
in the directions of vectors $\vac a$ and $\vac b$ and then $\vac a$ and
$\vac c$ is
        \begin{eqnarray}
        C(\vac a,\vac b)-C(\vac a,\vac c)
        &=&-\int d\lambda\ \!\wp(\lambda)\left[
        A(\lambda,\vac a)A(\lambda,\vac b)
        -A(\lambda,\vac a)A(\lambda,\vac c)\right]              \nonumber\\
        &=&-\int d\lambda\ \!\wp(\lambda)A(\lambda,\vac a)A(\lambda,\vac b)
        \left[1-A^{-1}(\lambda,\vac b)A(\lambda,\vac c)\right]  \nonumber\\
        &=&-\int d\lambda\ \!\wp(\lambda)A(\lambda,\vac a)A(\lambda,\vac b)
        \left[1-A(\lambda,\vac b)A(\lambda,\vac c)\right],       
	\end{eqnarray}
and, since $A(\lambda,\vac a)=\pm 1$,
        \begin{eqnarray}
        |C(\vac a,\vac b)-C(\vac a,\vac c)|
        &\le& \int d\lambda\ \!\wp(\lambda)|A(\lambda,\vac a)A(\lambda,\vac b)|
        \left[1-A(\lambda,\vac b)A(\lambda,\vac c)\right]       \nonumber\\
        &\le&\int d\lambda\ \!\wp(\lambda)
        \left[1-A(\lambda,\vac b)A(\lambda,\vac c)\right]       \nonumber\\
        &\le&\int d\lambda\ \!\wp(\lambda)
        \left[1+A(\lambda,\vac b)B(\lambda,\vac c)\right].      
        \end{eqnarray}
Any local hidden-variable theory should then satisfy the so-called Bell
inequalities:
        \begin{equation}
        |C(\vac a,\vac b)-C(\vac a,\vac c)|\le 1+C(\vac b,\vac c).
	\label{IneqBell}
        \end{equation}
for any set of vectors $\vac a,\vac b$ and $\vac c$. In contradistinction,
these inequalities are violated by quantum mechanics as one can check
by inserting the quantum correlation  (\ref{VarC9}) 
into (\ref{IneqBell}).
The experimental test of the Bell inequalities (see next section) allows to 
decide unambiguously
in favour of either quantum mechanics or local hidden-variable classical
theory. Note that
we have restricted our attention to local theories thus
non-local hidden-variable theories may violate the Bell inequalities too.

\subsection{A simple illustration of Bell's inequalities}
\begin{center}
\begin{figure}[!ht]
	\centerline{\hfill
	\psfig{figure=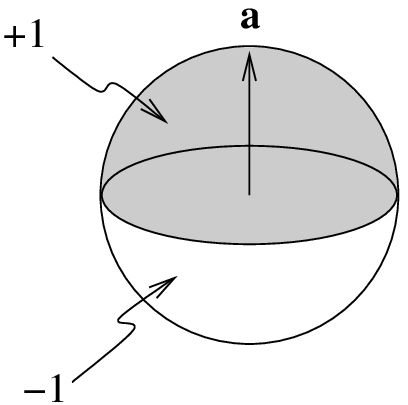,height=4cm}\hfill
	\psfig{figure=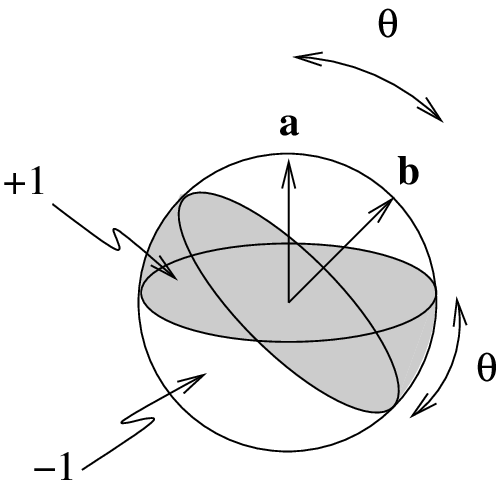,height=4.5cm}\hfill
	\psfig{figure=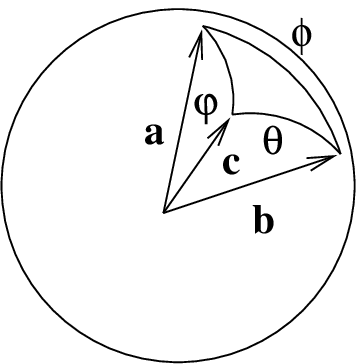,height=3.1cm}\hfill}
\caption{Illustration of the measurement in the hidden-variable theory considered in the text.}
\label{fig1}
\end{figure}
\end{center}

As a simple illustration, 
consider the following local hidden-variable theory.  Spins
are unit vectors and the measurement of $\bsigma.\vac a$
gives in a deterministic way the sign of $\bsigma.\vac a$.
Equivalently, the state of a spin can be associated with a point on the
unit sphere and the measurement of $\bsigma.\vac a$ gives $+1$ if this
point is in the hemisphere whose revolution axis is along $\vac a$ and $-1$
otherwise (see leftmost illustration of the figure~\ref{fig1}). An entangled pair is supposed to
be made of two spins in opposite directions so that $\bsigma_2=-\bsigma_1$.
The measurement of $\big(\bsigma_1.\vac a\big)\big(\bsigma_2.\vac b\big)
=-\big(\bsigma_1.\vac a\big)\big(\bsigma_1.\vac b\big)$ gives $+1$
if $\bsigma_1$ is only in one of the two hemispheres whose revolutions
axis are respectively along $\vac a$ and $\vac b$, and $-1$ otherwise.
As can be seen in the central diagram of figure~\ref{fig1}, the total surface giving an
output $+1$ is $4\theta$ where $\theta$ is the angle between the two
vectors $\vac a$ and $\vac b$. When assuming that the spins are produced in
any direction $\bsigma_1=-\bsigma_2$, the average correlation is
	\begin{equation}
	C(\vac a,\vac b)={1\over 4\pi}\big[+1\times 4\theta
	+(-1)\times 4(\pi-\theta)\big]={2\theta\over\pi}-1.
	\end{equation}
The Bell inequalities are satisfied:
	\begin{equation}
	|C(\vac a,\vac b)-C(\vac a,\vac c)|={2\over\pi}\big|\theta-\varphi|
	\le {2\phi\over\pi},
	\end{equation}
where $\varphi$ is the angle between $\vac a$ and $\vac c$ and $\phi$ between
$\vac b$ and $\vac c$. Note that the angles $\theta$, $\varphi$ and
$\phi$ are the length of the three edges of the triangle
on the unit sphere joining the vectors $\vac a$, $\vac b$ and $\vac c$
(see the rightmost illustration of  figure~\ref{fig1}).
The equality is obtained when the three vectors are in the same plane.

\section{Tests of Bell's inequalities: from thought experiments to
realistic experiments.}

The conclusion of the above paragraph is that the Bell
inequalities provide a quantitative criterion to test the
hidden-variable theories versus quantum mechanics. Bell's inequalities
brought the discussion into the experimental domain. Many experiments 
have since been performed, the outcomes of which are 
consistent with quantum mechanics and inconsistent with local
realism. 

The first step through experimental realization was performed
in 1969 by Clauser, Horne, Shimony and Holt~\cite{ClauserEtAl1969}. They 
generalised the Bell theorem so as to apply to actual
experiments (CHSH inequalities). At the end of their paper, they
proposed an experiment based on the Bohm and Aharonov~\cite{BohmAharonov1957} 
version
of the EPR paradox. It is an optical variant aimed at
measuring the polarisation correlation of a pair of optical photons.
This experiment was the first of a long series in quantum optics.

The two main assumptions
underlying the formalism leading to Bell's theorem are:
\begin{description}
\item{i)} Existence of hidden variables which
renders an account of the correlations. It is a 
necessary hypothesis to obtain a conflict between the Bell inequalities and
quantum mechanics.
\item{ii)} Locality: the result of
a measurement of the polarisation of one of the photons does not depend on
the orientation of the other analyser
and vice-versa and the way the pairs are emitted by the source does
not depend on the measurement setup. The locality assumption
is thus crucial: Bell's inequalities will no longer hold without
it.
\end{description}

Let us stress that determinism does not seem to be a necessary 
ingredient, since 
it is possible to
obtain analogous inequalities 
with local stochastic 
hidden-variable theories~\cite{TheseAspect,DEspagnat1971}.
Indeed, in 1969 Bell himself  gave an example where the local hidden
variables do not determine completely the outcome of the
measurements. One can imagine for example
that, for whatever reason, some local fluctuations occur
and are responsible for the stochastic character of the 
measurement outcomes.
However such a theory is still in conflict with quantum mechanics.

\subsection{Optical variant of the Bohm experiment}
\begin{figure}
%
\vskip-0mm
\hskip-10mm\includegraphics [scale=0.5]{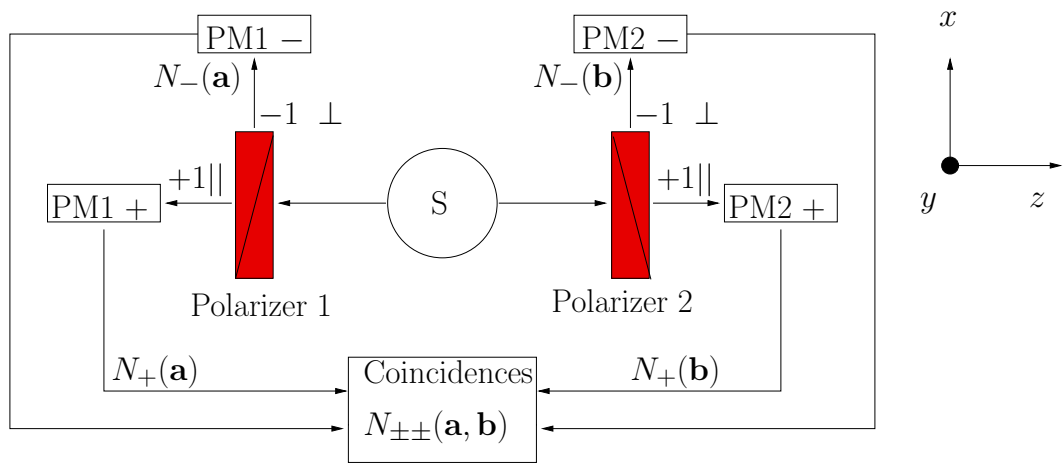}
\vskip-70mm
%
\caption{
Optical version of the Bohm experiment. S is the source of
photons. PM are photo-multipliers.}
\label{FigCathy1}
\end{figure}

Figure~\ref{FigCathy1} illustrates the optical variant~\cite{Aspect1999} 
of the Bohm version~\cite{BohmAharonov1957} of the EPR
thought experiment.
The source $S$ produces pairs of photons with different frequencies
$\nu_{1}$ and $\nu_{2}$, sent in opposite directions. 
Each photon encounters an analyser (a polariser able to
measure the linear polarisation) whose orientation can be set by
the experimentalist. The emerging signals are detected by
photo-multipliers along two perpendicular directions. A 
coincidence monitor  measures
the probabilities of single or joint detections. The analyser I
(respectively analyser II) is oriented along the \textbf{a}
(respectively \textbf{b}) direction. The result of the measurement
of photon $\nu_{1}$ is $A({\bf a}) =+1$ if the polarisation is
parallel to \textbf{a} and $-1$ if the polarisation is perpendicular to
\textbf{a}. 
The result of the measurement of photon $\nu_{2}$ is
$B({\bf b}) =+1$  if the polarisation is parallel to \textbf{b} and $-1$
if the polarisation is perpendicular to \textbf{b}.

The polarisation part of the state vector describing the pair is
an entangled state:
\begin{equation}
\label{eq1}
\left| {\psi (1,2)} \right\rangle =\frac{1}{\sqrt 2 }\left[ {\left| {x,x}
\right\rangle +\left| {y,y} \right\rangle } \right],\label{eq18}
\end{equation}
where $\left| x \right\rangle $ and $\left| y \right\rangle $ are
one-photon linear polarisations states. 
The polarisation correlation coefficient for polarisers in orientations
\textbf{a} and \textbf{b} is
\begin{eqnarray}
\label{eq2} C({\bf a},{\bf b})&=&[N_{++} ({\bf a},{\bf b})+N_{--}
({\bf a},{\bf b})-N_{-+} ({\bf a},{\bf b})-N_{+-} ({\bf a},{\bf b})]/N 
\nonumber\\
&=&P_{++} ({\bf a},{\bf b})+P_{--} ({\bf a},{\bf b})-P_{-+} ({\bf
a},{\bf b})-P_{+-} ({\bf a},{\bf b}),
\end{eqnarray}
where $N$ is the rate of emission of pairs, $N_{\pm \pm
}$(\textbf{a},\textbf{b}) are the coincidence rates and $P_{\pm \pm
}$(\textbf{a},\textbf{b}) the empirical
probabilities of joint detection.
Using the entangled state (\ref{eq18}),
the Bell inequality reads as
        \begin{equation}
        |C(\vac a,\vac b)-C(\vac a,\vac c)|\le 1-C(\vac b,\vac c).
	\label{IneqBellPhotons}
        \end{equation}

\subsection{Towards experimental tests: the CHSH inequalities. }

\begin{figure}
\begin{center}
\vskip-10mm
\hskip-35mm\includegraphics [scale=0.4]{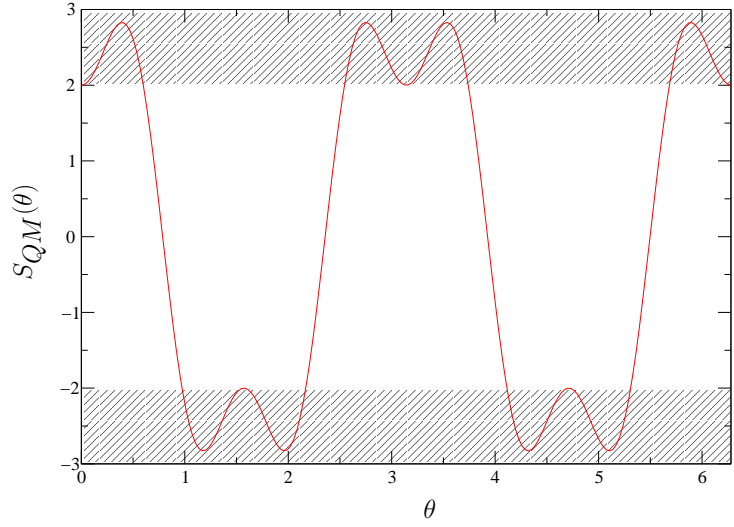}
\end{center} 
\vskip-45mm
\caption{
S$_{QM}$ versus $\theta $ as predicted by quantum mechanics for
EPR pairs. The conflict with Bell inequalities happens when
$\vert $S$\vert $ is larger than 2.}
\label{FigCathy2}
\end{figure}

In 1969, Clauser, Horne, Shimony and Holt~\cite{ClauserEtAl1969} 
expressed the Bell
inequalities in terms of experimental quantities, namely coincidence
rates measured for 4 directions ${\bf a}$, ${\bf a}'$, ${\bf b}$,
and ${\bf b}'$ of the polarisers,
\begin{eqnarray}
\label{eq3} -2\le S\le 2 \mbox{ with  }
  S=\left| {C({\bf a},{\bf b})-C({\bf a},{\bf b}')}
\right|+C({\bf a}',{\bf b})+C({\bf a}',{\bf b}').\label{eqS}
\end{eqnarray}
Using  quantum mechanics,  the probabilities of detections
are
\begin{eqnarray}
\label{eq5} P_{+} ({\bf a})&=&P_{+} ({\bf b})= P_{-} ({\bf a})=P_{-}
({\bf b})=\frac {1} {2}\nonumber\\
P_{++} ({\bf a}, {\bf b})&=&P_{--} ({\bf a},{\bf b})= \frac{1}{2}
\cos ^{2} ({\bf a},{\bf b})\nonumber\\
 P_{+-}({\bf a}, {\bf b})&=&P_{-+} ({\bf a},{\bf b})= \frac{1}{2}\sin ^{2}
({\bf a},{\bf b}),
\label{eq3bis}
\end{eqnarray}
leading to 
the correlation coefficient 
\begin{eqnarray}
C_{QM} ({\bf a},{\bf b})=\cos 2 ({\bf a},{\bf b}).\label{eqCQM}
\end{eqnarray}
The quantum mechanical expression for $S$  is thus
\begin{equation}
S_{QM} ({\bf a},{\bf b})=\cos 2 \theta+\cos 2 \theta'+\cos 2
\theta''-\cos 2 (\theta+\theta'+\theta''),\label{eqSQM}
\end{equation}
with $(\textbf{a},\textbf{b})=\theta $,
$(\textbf{b},\textbf{a}')=\theta'$,
$(\textbf{a}',\textbf{b}')=\theta'' $, $(\textbf{a},\textbf{b}')=
\theta + \theta' + \theta'' $.
The variation of $S_{QM}$ when $\theta =\theta'=\theta'' $ is
shown in figure~\ref{FigCathy2}.
The conflicts with CHSH (\ref{eqS})
inequalities occur when the absolute value of $|S|$ is larger than 
$2$ ($\mathop S\nolimits_{QM}^{\max } =2\sqrt 2$ and
$\mathop S\nolimits_{QM}^{\min } =-2\sqrt 2$).
However, it is clear that there are many orientations for which
there is no conflict.

However, in the 70's, two-channel polarisers were not available
and experiments were performed with one channel polarisers. Such
polarisers transmit light polarised parallel to \textbf{a} (or
\textbf{b}), but not the orthogonal counterpart so that only the
$N_{+}$(\textbf{a}), $N_{+}$(\textbf{b}) counting rates and the
$N_{++}$(\textbf{a},\textbf{b}) coincidence rate are measured. The
above Bell inequalities have to be formulated again in order to
depend only on these measured quantities.
Moreover, in realistic experiments only 10$^{-5}$  of the emitted photons are
detected, so the quantities involved in $S$ are not of the same order of
magnitude and there is no more possibility of violation of the generalised
Bell inequalities. In fact, one needs to consider only the pairs actually
detected. To do so, it is necessary to measure the rates when polarisers
are removed. The final version of the CHSH inequality is then
\begin{eqnarray}
\label{eq4} 0\le S'\le 1,
\end{eqnarray}
with
\begin{eqnarray}S'&=&\frac{1}{N(\infty ,\infty )}
\Bigl[ N_{++} ({\bf a},{\bf b})-N_{++} ({\bf a},{\bf b'})+ N_{++}
({\bf a'},{\bf b})\nonumber\\
&&+N_{++} ({\bf a'},{\bf b'})-N_{++} ({\bf
a'},\infty )-N_{++} (\infty ,{\bf b}) \Bigr].
\end{eqnarray}
The symbol $\infty$ conventionally indicates the absence of a polariser.
The important point is that $S'$ now depends only on measured quantities of
the same order of magnitude.
For single channel polarisers, $\mathop S\nolimits_{QM}^{'\max } =0.207$ and
$\mathop S\nolimits_{QM}^{'\min } =-1.207$ correspond to the maximum conflict.

Experimental inefficiencies (polariser defects including non-perfect
transmission rates, non-finite detection angle, accidental
birefringence, etc.) lead to a decrease of $C(\textbf{a},\textbf{b})$.
$S_{QM}(\textbf{a},\textbf{a}',\textbf{b},\textbf{b}')$ or
$S'_{QM}(\textbf{a},\textbf{a}',\textbf{b},\textbf{b}')$ is then reduced,
hence eventually the conflict with Bell's inequalities may be undetectable.
Actual transmission coefficients and finite
solid angles lead to $\mathop S\nolimits_{QM}^{'\max } =0.06$ and $\mathop
S\nolimits_{QM}^{'\min } =-1.06$. A sensitive test with one channel
polarisers is thus possible if the actual experiment is as close
as possible to the thought experiment.

\subsection{First experiments in the 70's.}
Following the CHSH paper, a first series of tests was
performed in the 70's using polarisation-correlated visible photons. The
entangled pair is produced by atomic radiative cascades. A beam of
atoms is emitted by an oven in a vacuum chamber and submitted to an
excitation. During the de-excitation process, two photons
are produced nearly simultaneously if the lifetime of the
intermediate level is short enough. For example, for the
0-1-0 calcium cascade ($4p^{2}$ $ ^{1}S_{0}\to 4s 4p
$ $^{1}P_{1} \to 4s^{2} $ $^{1}S_{0}$) a green and a
violet photon are produced and the lifetime of the intermediate
level is very short, namely 5ns.

A first experiment was performed by Clauser and Freedman in
Berkeley in 1972~\cite{FreedmanClauser1972}. 
They use the 0-1-0 Ca cascade described
above. The excited state was obtained by ultraviolet radiation. In
fact, additional spurious cascades occur and, because of
consequently weak signal, the experiment took 200~hours. A
violation of Bell's inequalities by 5 standard deviations was
observed.
Experiments have also been performed using a 1-1-0 cascade of
mercury. The first one was in disagreement with quantum 
mechanics~\cite{Pipkin1978}. 
However,  the experiment reproduced by another group leads to
a significant violation of Bell's inequalities~\cite{Clauser1976}. This last
experiment took 412~hours.
Finally, in 1976, Fry and Thompson~\cite{FryThompson1976} produce 
entangled pairs with a rate of several orders of magnitude larger.
To obtain such a gain, they excited the
upper level \textit{selectively} using a laser. The data was
collected in 80~minutes and was in excellent agreement with
quantum mechanics, within 4
standard deviations.

Thus, at the end of the 70's, the experiments were already in agreement
with quantum mechanics. However, they were performed with single channel
polarisers  and one has to keep in mind that their
use requires indirect reasoning, and auxiliary calibrations.

\subsection{Experiments of the Orsay group (1981-1982)}
Thanks to progress in laser physics and modern optics, a second generation
of experiments was carried out by the Orsay group in the early 80's.
They developed a high efficiency stable and well controlled source of
entangled photons. Using the same calcium cascade as 
Freedman \textit{et al},
Aspect \textit{et al} performed {\em direct} excitation using two lasers,
so that
the atom radiative
decay delivered \textit{only} pairs of entangled photons.

The first experiment  was  performed
using single channel polarisers~\cite{AspectEtAl1981},  with an 
improved 
analyser\footnote{Each analyser was 
a pile of plate polarisers.} ensuring excellent rotational
invariance and good transmission coefficients. 
Moreover, the main
difference with previous experiments was that the excitation rate 
was more than ten times
larger, allowing 
a large
variety of tests to be performed with a good statistical accuracy. 
Firstly they performed measurements for the full 360$^\circ$ 
range of relative orientation of the polarisers.
The experimental value of $S'$ was found equal to 0.126$\pm $0.014 violating
the Bell inequalities by 9 standard deviations and in excellent agreement
with the value calculated from quantum mechanics including
the polarisers efficiencies and the lenses apertures angles, 0.118$\pm
$0.005.
Secondly, no change was observed in the results 
with polarisers at a distance (6 m)
larger than the coherence length of $\nu_{2}$ (1.5~m).

The second experiment of the Orsay group, using two-channel
polarisers\footnote{The polarisers were polarising cubes, made of two 
prisms with
suitable dielectric thin films on their sides stuck 
together.}~\cite{AspectEtAl1982a} 
was  the first actual analog
of a Stern-Gerlach filter for spin $\raise.5ex\hbox{$\scriptstyle
1$}\kern-.1em/ \kern-.15em\lower.25ex\hbox{$\scriptstyle 2$} $
particles. 
Five runs were performed for each sensitive orientation of the
polarisers. The average yielded $S= 2.697\pm 0.015$ in very
good agreement with quantum mechanics ($S_{QM}=2.70 \pm 0.05$).
This result leads to the greatest violation of generalised Bell
inequalities ever achieved (40 standard deviations!).

All the above experiments were static experiments: the locality
condition -- which is crucial -- is indirectly tested. In 1964,
J.S Bell already noted that ``{\em the setting of the instruments are made
sufficiently in advance to allow them to reach some mutual rapport
by exchange of signals with velocity less than or equal to that of
light}''. Consequently, the important point was to perform an experiment
in which the settings are changed during the flight of the
particles. The locality condition would then be compatible with
Einstein causality preventing any faster-than-light influence.
Such a timing experiment was done by the Orsay group in 
1982~\cite{AspectEtAl1982b}. 
During this third experiment, the setup was modified 
in order to switch the direction of polarisation analysis after the photons 
left the
source. To do
so, the (single-channel) polarisers were replaced by a switching
device together with two polarisers in two different orientations, 
equivalent to a variable
polariser switched between  two polarisations. 
The switchings are periodic, but the relative phases
were randomly chosen and uncorrelated. 
The distance between the switches was
large enough (13 m) for the
time of travel of a signal at the velocity 
of light (43
ns) to be significantly larger than the delay between two switchings (about 10
ns) and the delay between the emission of the two photons (5ns
average).

Because of a reduced signal (due to the limited aperture of the switches), a
15~hours averaging was necessary. The results are in good agreement with
quantum mechanics, violating the Bell inequalities by 5 standard
deviations.

\subsection{Toward the ideal experiment}
After the above work of
the Orsay group in the early 80's, one could  conclude that
all recent experiments confirm the predictions of quantum
mechanics and that no additional test is required.
However, from a strictly logical point of view, the above experiments
do not succeed in ruling out a local realistic explanation
\textit{completely}, because of two essential loopholes: 
locality and detection.

The locality loophole remains open
in the third experiment by Aspect \textit{et al.}~\cite{AspectEtAl1982b} 
due to the periodic sinusoidal switching used.
Thus communication slower than the
speed of light, or even at the speed of light, could in principle
explain the results obtained. 
In fact, the condition of locality is \textit{fully enforced}
for the first time only in 1998 by Weihs and co-worker~\cite{WeihsEtAl1998} in
Innsbr\"{u}ck. In their experiment, the necessary space-like
separation of the observations is achieved by:
\begin{description}
\item{i)} sufficient physical distance between the measurement stations (440m),
\item{ii)} ultra-fast and random setting of the analysers,
\item{iii)} completely 
independent data registration, the timing being monitored by
local rubidium atomic clocks.
\end{description}
This crucial experiment has become possible because of the
development of new sources of correlated photons~\cite{KwiatEtAl1995}. In these
sources, a pair of red photons is produced by a parametric down
conversion of a U.V. photon (pump beam) in a non-linear crystal.
Because of the phase matching condition (determined by the
orientation of the crystal axes relative to the pump beam), there
is a strong correlation between the directions of emission of the
two photons of a pair (in contrast with the atomic
radiative cascades which produce photons only weakly correlated in
direction). Consequently, larger coincidence rates can be
obtained (more than one order of magnitude larger than in atomic
radiative cascade). Moreover, the production of two narrow beams
of correlated photons permits one to feed them into two optical
fibres. Then, from a practical point of view, with this new
schemes:
\begin{description}
\item{i)} it becomes 
possible to work with small integrated electro-optical devices,
\item{ii)} the detectors can be kilometres apart.
\end{description}
Weihs \textit{et al} observed a 
violation of the CHSH inequality of 100 standard
deviations! These results close  the locality loophole.

It is worth noting that the new parametric down conversion sources
present another interesting feature. They produce entangled states
with correlation between other observables  than polarisation, namely
time of emission, energy, direction of emissions. Among the amount
of experiments performed during the 15 last years, two of them can
be mentioned: they present an experimental demonstration of
quantum correlations over 4km~\cite{TapsterEtAl1994} 
and 10km~\cite{TittelEtAl1998} using optical
fibres, concluding  that the distance does not destroy the
entanglement.

The  so-called ``detection loophole''
relies on the fact that all experiments so far detected only a small
subset of all pairs created~\cite{Grangier2001}. It is therefore necessary to
assume that the pairs registered are a fair sample of all pairs
emitted (``fair sampling assumption") as emphasised by
J.S. Bell: ``\dots\textit{ it is hard for me to believe
that quantum mechanics works so nicely for inefficient practical
set-ups and is yet going to fail badly when sufficient refinements
are made''.} However, experiments have been performed in order to
try to close the detection loophole.
Due to a too low
efficiency of currently available photon detectors, 
people also used massive particles which
are easier to detect. Conclusive tests of the Bell
inequalities 
have been realised by Rowe \textit{et al}~\cite{RoweEtAl2001}. Two
trapped ions are
prepared in an entangled state by Raman laser beams. The
agreement with quantum mechanics is excellent, making their
experiment the first violation of the Bell inequalities with high
enough efficiency.
However one has to mention 
that the two ions in the same trap are very close to each
other, so that the detection events are not space-like separated and it
seems difficult to fulfil the timing conditions in experiments following
this scheme.

Entanglement experiments have also been performed with Rydberg
atoms and microwave photons in a cavity. However, the signal
contrast in these experiments~\cite{RaimondEtAl2001} 
is not high enough to observe a
violation of the Bell inequalities.

\section{Neo-Bohmian interpretation}
\subsection{Orthodox interpretation, a few comments}
In the physics community it is frequently stated that ``it is not possible 
to know simultaneously the position and the momentum
of a given particle, since the position and momentum 
operators are non-commutative". 
This statement leads to a naive realistic point of 
view concerning the use of operators in quantum 
mechanics, as primary ``real" objects and not as 
emerging quantities. Let us say, with Bohr, that 
what we call ``quantum observables"  
acquire their significance only through their 
association with specific experiments (measurements).
In the Hilbert space parlance, the relevant quantities 
associated with a given experiment are the partition of subspaces 
${\cal H}_{\alpha}$ and the values $\lambda_{\alpha}$ 
associated with the final result of the experiment. The collection
$\{{\cal H}_{\alpha},\lambda_{\alpha}\}$ is compactly 
represented by the self-adjoint operator 
\be
A=\sum_\alpha \lambda_\alpha P_\alpha\; ,
\ee
where $P_\alpha$ is the projector onto the corresponding 
Hilbert subspace ${\cal H}_{\alpha}$.
In this way a given experiment $\epsilon$ is 
associated with a given operator $A_\epsilon$ which 
finally is interpreted as 
having a reality that is linked to the system 
alone and independent of the actual reproducible experiment.

\subsection{Bohmian interpretation}
\subsubsec{The goals of such an interpretation}
The principal goal of the Bohmian interpretation is clear -- 
to restore realism. It is an attempt 
to give a clear ontology to the constitutive elements of the 
world~\cite{deBroglie1927,Bhom1952,Durr1992}.  
On the most simple level of quantum mechanics, we can state within this 
interpretation that the fundamental elements are particles which are 
really particles and consequently they have a perfectly definite 
position and velocity at all times.  This does not mean that such an 
interpretation is focused on the particle like structure of matter, since 
a field ontology is also perfectly possible. Once more, the goal is to 
give a clear ontology. 
Within the particle picture, the lack of causality/determinism 
of quantum mechanics 
is of the same type as that encountered in classical statistical mechanics, 
basically due to our lack of knowledge of the initial conditions.
But one should not confuse this approach with a return to the classical 
Newtonian concepts. 

\subsubsec{de Broglie - Bohmian theory}
In 1927, at the Solvay congress, the  young French 
physicist Louis de Broglie proposed 
a quantum theory known as the pilot-wave theory~\cite{deBroglie1927}.  
His theory is grounded on two hypotheses: 
\begin{description}
\item{i)} there exists a 
continuous field $\phi=Re^{iS/\hbar}$ satisfying 
the Schr\"odinger equation and  
\item{ii)} the necessity 
that a particle, of mass $m$, follows the 
trajectories defined by $m{\bf v}=\nabla S$.
\end{description}
This is basically the interpretation proposed 
some thirty years later by David Bohm~\cite{Bohm1952}.  
The complete description 
within the pilot-wave interpretation of a system 
of $N$ particles is specified by the wave function 
\be
\Psi(q,t)\quad q=(q_1,q_2,\dots,q_N) \in {\field{R}}^{3N},
\ee
and the coordinates,
\be
Q=(Q_1,Q_2,\dots,Q_N) \in  {\field{R}}^{3N},
\ee
of the particles positions. 
The wave function $\Psi(q,t)$ evolves according to the Schr\"o\-din\-ger equation
\be
i\hbar \frac{\partial }{\partial t}\Psi= H \Psi,
\ee
where 
\be
H=-\sum_k \frac{\hbar^2}{2m_k} \nabla_k^2 +V,
\ee
and pilots the motion of the particles according to the equation
\be
\frac{{\rm d}Q_k}{{\rm d}t}=\frac{\hbar}{m_k} \left. \frac{{\rm Im} 
(\Psi^*\nabla_k\Psi)}{\Psi^*\Psi}\right|_{Q_1,\dots,Q_N}\; .
\ee

From here, one can try to write a generalised Jacobi 
equation where, together with the classical potential, a ``quantum potential" 
appears, leading to a modification of Newtonian mechanics via a quantum 
force. This attempt however is not really suitable since it does not 
emphasise 
the profound epistemological change from Newtonian mechanics to quantum 
mechanics. 
Putting the wave function 
$\Psi=R e^{iS/\hbar}$ into 
the Schr\"odinger equation,
the real and imaginary 
part lead to two equations. the first of these is  the Hamilton-Jacobi equation
\be
\frac{\partial S}{\partial t} + H(\nabla S,q) +U=0,
\ee
where 
\be
U=-\sum_k\frac{\hbar^2}{2m_k}\frac{\nabla_k^2 R}{R}
\ee 
is the so-called quantum potential, the 
only quantity in the Hamilton-Jacobi 
equation proportional to the Planck constant~\cite{deBroglie1927,Bohm1952}. 
The second equation is a continuity equation,
\be
\frac{\partial \rho}{\partial t} + {\rm div} \left(\rho v\right)=0,
\ee
with $v=(v_1,\dots,v_N)$ and $\rm div$ the 
divergence in configuration space. The particle velocities are given by
\be
v_k=\frac{\nabla_k S}{m_k}\; .
\ee
This rewriting of the quantum equations 
suggests that the quantum nature of 
matter lies only on a slight modification 
of the classical equations, basically by the 
appearance of the quantum potential. 
This 
interpretation is tempting and was indeed 
adopted in the past. However the de Broglie-Bohm theory 
should  actually be considered as a new theory with its own corpus of 
concepts~\cite{Durr1997}. 
For example, the particle masses are not 
the coefficients appearing in Newton dynamical 
law but rather the factors of the field equation 
and entering into the guiding equations. On the 
same lines, the velocities in the  de Broglie-Bohm 
theory are not independent of the positions since 
they are given by the guiding equations. What is 
fundamental here in this interpretation of the theory 
is the guidance condition and not the quantum potential
which can be viewed at best as a good picture when 
taking the classical limit of the theory.

From the very beginning, the pilot-wave theory had to face several 
criticisms. The first is  that there is no back reaction of the particles 
on the field $\Psi$, which is unusual in physics. 
Another criticism is that the field itself, contrary to a physical field, 
doesn't live in the physical space but rather in the much larger 
configuration space ${\field{R}}^{3N}$. 
This leads to the non-local and non-separable 
character of the theory. Finally, 
if one requires that this theory is consistent  with the usual
quantum theory predictions, one 
should demand that the positions of the particles 
at the initial time $t_0$ should be distributed according to the law
\be
\rho(q,t_0)=|\Psi(q,t_0)|^2,
\ee
and they will be distributed at a latter time as
\be
\rho(q,t)=|\Psi(q,t)|^2\; .
\ee
One of the criticisms was that within this theory, the wave 
function plays at the same time two apparently 
irreconcilable roles. It determines the quantum potential 
acting on the particles and  define a probability 
density associated with the trajectories. 
Evidently, the most important point is to give a 
strong argument, possibly of the same type as that 
given in classical statistical mechanics, leading to 
the quantum distribution of the initial positions~\cite{Bohm1953}.

\subsubsec{Probabilities}
The important 
point in order to recover and save the usual quantum 
mechanics power is that the probability distribution 
for a system, which has an associated wave 
function $\Psi$, should be given by
$
\rho(q)=|\Psi(q)|^2
$
at an initial instant. If this is so, the 
continuity equation, extracted from the Schr\"odinger equation,
will guaranty that the system will 
stay $|\Psi(q,t)|^2$ distributed at a later time.  
This condition is called the quantum equilibrium 
hypothesis~\cite{Bohm1953,Durr1992}.  
Accordingly, on a system in quantum equilibrium
it is not possible to go beyond the distribution $|\Psi|^2$ and 
consequently no deviation or violation of the  Heisenberg 
inequalities will be ever observed. In other words, 
the clear ontology introduced via, for example, the 
particle picture, will neither remove nor erase 
the ``paradoxical" quantum behaviour of the system, 
since it is in quantum equilibrium. The gain compared 
to the orthodox interpretation is that now the 
indeterminism is empirical not ontological. 

The remaining point is to address the following 
question. Why is the system of interest in quantum equilibrium?
From the very beginning of the de Broglie-Bohm interpretation, 
this was the difficult aspect to be clarified. Several attempts were 
made in several directions, modifying the original 
theory or trying to set up a 
dynamical foundation 
of the quantum equilibrium hypothesis. 
It is not clear  if such an attempt is accessible or feasible. 
However it could be as well that 
the equilibrium hypothesis has to be postulated
as an empirical fact. 
More recently, in another direction of thought, D\"urr {\it et al.} 
proposed an explanation {\it \`a la } Boltzmann~\cite{Durr1992}. 
The essential point is the notion of a partial wave function associated 
with a subsystem of a much larger system, exactly in analogy with 
the emergence of the canonical distribution when 
considering a small part of 
an isolated total system, the total system being in a ``typical" 
state. 
Indeed, if one takes seriously enough the de Broglie-Bohm theory one should 
expect that the behaviour of a subsystem is completely determined by the 
wave function $\Psi(q)$ of the universe and its corresponding configuration. 
D\"urr {\it et al.} associate a wave function with a subsystem within the 
following decomposition of the universe wave function,
\be
\Psi(x,y)=\psi(x)\Phi(y)+\Psi^{\perp}(x,y)
\ee
where $x$ stands for the system variable while $y$ is associated 
with the rest of the universe. The functions $\Phi$ and $\Psi^{\perp}$ 
have macroscopically distinct supports for the variable $y$. So, if the 
configuration variable $Y$ lies into the $\Phi$ support, we have
\be
\Psi(x,Y)=\psi(x)\Phi(Y),
\ee
and one is able to associate a wave function with the subsystem. The 
preceding structure of the total wave function is justified within the 
decoherence mechanism, {\it i.e.} the irreversible flow of the coherent 
phase into the so called environment, that is the rest of the universe. 

Since, as stated by Mach and others, the universe is given only once, it 
is meaningless to associate with it a distribution $\rho$ from its wave 
function $\Psi$. Nevertheless, D\"urr {\it et al.} argued on the base 
of ``time independence'' that $|\Psi|^2$ gives a measure of typicality and 
then,  within a typical universe that is over an 
overwhelming majority of possible initial universe configurations, the 
equilibrium hypothesis holds for the subsystems\cite{Durr1992}. 
Objections to this derivation were formulated since it is not clear at 
all why dynamical considerations should play any role 
in order to specify the 
initial conditions. 

\subsubsec{The wave function as a law}
In their formulation of Bohmian mechanics, D\"urr {\it et al.}  proposed 
a very interesting interpretation of the wave function as a law governing 
the dynamics rather than specifying the state of the system~\cite{Durr1997}.  
Indeed, if we suppose that the function of the universe is given by a law 
of the form
\be
H_U \Psi=0\; ,
\ee
the solution $\Psi$ of this equation together with the initial configuration 
$Q$ determines perfectly the future evolution of the positions, thanks to 
the guiding equation,
\be
\frac{\rm d}{{\rm d}t}Q= D (\log \Psi),
\ee
where $D$ is a differential operator.
We see here the analogy with, for example, Hamilton's equations 
\be
\frac{\rm d}{{\rm d}t} \xi= D H
\ee
and one is tempted to identify the role of $\log \Psi $ with that of the 
Hamiltonian $H$. In that sense, $\Psi$ plays the role of the law 
governing the motion of the particles. This interpretation has also the 
merit of providing an explanation of the ``bizarre" fact that the wave 
function lives in the configuration space and not in the real space 
as other physical fields.  

Finally, within this picture, one may notice that the time-dependent 
Schr\"odinger equation for a system should emerge from the universe 
equation $H_U \Psi$ when restricting to the system wave function $\psi$. 
In sufficiently simple models such a reduction was accomplished explicitly.

\subsection{Criticism of the de Broglie-Bohm theory}
Since the very beginning, what is usually called the de Broglie-Bohm 
theory or the pilot-wave theory has received much criticism.
These include the return to classical concepts, the lack of 
new predictions compared to the standard quantum mechanics, asymmetry 
with respect to position and momentum, the introduction of myriads of 
empty waves, the foundation of the quantum equilibrium hypothesis, 
as well as its relativistic and quantum field generalisations. 
This list is certainly not exhaustive but it is intended to be long enough to give to the 
reader the impression of a not-so-well founded theory. However, most 
of the criticism could be answered with convincing arguments and none
of these objections are rigorous disproof 
of the de Broglie-Bohm theory. One can refer to Passon~\cite{Passon2004} 
for a short reply to 
all these objections. 

\section{Discussion}
Realism is the assumption that there exists an objective external world  
independent of our perception of it.
In a realistic physical theory, one thus requires 
a clear ontology of the basic ``objects'' used -- for example fields
which are really fields, particles which are really 
particles, etc.~\cite{Durr1992,Durr1997}.
Locality means that these objects
are defined locally with no instantaneous action at a distance. 
Local realism may thus be defined by the combination of the principle of 
locality with the assumption that all objects must objectively have 
their properties already before they are observed.
The paradigmatic example is that of local
hidden variables. 

To summarise the experimental 
contribution to this debate, one can conclude that, 
since the pioneering ones in the sixties, an impressive 
amount of studies are in agreement with quantum mechanics, 
or at least contradict local hidden-variable theories.
Today, even in the absence of an ultimate experiment - where
both the detection loophole would be 
definitely closed \textit{and} 
locality enforced - one can conclude there is a failure of local
realism. 
It would be interesting to improve
the detection process in such experiments, as recently proposed by
Garcia-Patron \textit{et al }~\cite{Garcia-PatronEtAl2004}. However
we do not consider that the detection loophole is as crucial as
the locality one.

It is nevertheless important to underline that local 
hidden-variable theories have been
ruled out but that the hidden-variable theories in general have not been
disqualified. For example, the Bohmian mechanics
or  stochastic approaches 
are not invalidated, since 
they are non-local theories.
Although the neo-Bohmian interpretation, 
offers a promising alternative and gives an example 
of the possibility of a deterministic interpretation with a clear ontology,
it suffers from an essential weakness.  
This is the overly-close relationship with 
the usual quantum mechanics, since it was specially designed to 
restore the predictions made by ordinary quantum mechanics. 
Consequently, neo-Bohmian  
and Quantum Mechanics
are examples of empirically equivalent theories and
it is not clear whether one should expect a significant progress in the 
interpretation restricting only to 
empirically equivalent theories.
Stochastic approaches open a perspective, but suffer from the difficulty
of admiting a generalisation in a relativistic context.

Eventually we 
can summarise the basic lines of the orthodox interpretation as being 
within  the essential probabilism, 
forbidding the possibility of an ontological determinism,  
and the complementarity principle, stating that it is only possible 
within one experiment to reach partial aspects of reality 
which are excluded mutually.
Consistently with this approach, one may comment that 
the perfect knowledge of the state of the system leads in general, 
{\em nevertheless}, to probabilistic aspects, that is to 
{\em intrinsic stochasticity}.
The truth is not elsewhere, deeper, 
but on the surface. In some sense, we recover 
again the idea of an achievement of science, an old idea 
that seems to accompany the new theories  frequently. 
It is amazing to notice that so much efforts 
of research and pedagogical type have been done into the 
direction of impossibility theorems by those who 
themselves have discovered and built the theory. In other areas 
of human activity this would be suspicious, in physics it is not.

\section*{Acknowledgements}
We acknowledge Malte Henkel for his kind help in the translation of 
quotations from Einstein.
We also would like to express our gratitude to Ralph Kenna for his careful 
reading of the mansucript, and more precisely for his rewriting 
of many sentences, 
trying to make a ``frenglish'' text almost 
british, or at least readable by 
english-speaking persons.

\def\paperBB#1#2#3#4#5#6{#1, #2  {\bf #3}, p.\ #5 (#4), \ {\em #6} }
\def\bookBB#1#2#3{#1, {\em #2}, #3}

\section*{Questions and answers}
\begin{itemize}
\item[${\cal Q}$] {\em (Ara Apkarian)}: Could you 
comment on the conceptual distinction between locality and
separability.

\item[${\cal A}$] 
One may define separability in the following way: a separable state is one that can be created from another state by local means, which is not true for an entangled state. 
Mathematically, one expresses that in the form
$$
\rho=\sum_k p_k \rho_A^k\otimes\rho_B^k\; , \qquad \sum_kp_k=1
$$
where $\rho$ is the density matrix of the quantum state and where 
$\rho_{A,B}$ are density matrices belonging to the Hilbert spaces 
${\cal H}_{A,B}$.
It seems then that locality plays a significant role in the concept of separability. However, if one takes as a definition for locality the fact that distant objects cannot influence each other directly, 
one implicitly refers to objects, meaning real objects. But, in the so called orthodox interpretation, the wave function have no direct physical interpretation or reality, meaning that it is the concept of local object that is rejected in general. In a non separable state, one cannot a priori define local real objects.

\item[${\cal Q}$] {\em (Eduardo Lude\~na)}: Could you 
comment on the effect of distance on entanglement.

\item[${\cal A}$] What is important, rather than distance, is the interaction
with environment. As long as this interaction remains negligible, the 
entangled character of a quantum state is preserved.

\item[${\cal Q}$] {\em (Bertrand Berche)}: This is a personal 
comment more than a question. I believe that physicits' common conception
of a physical theory is very constraining. Could'nt we imagine as
admissible a theory which would not be able at all to answer 
(I mean even not in probabilistic terms) 
some questions which 
are {\em a priori} from its domain of applicability? 
Something a bit like G\"odel's theorem and propositions which are 
neither true nor false.
Why do we ask so much to our theories? Isn't it due to our custom 
that physical theories have been so powerful in the past, with the 
"The Unreasonable Effectiveness of Mathematics in the Natural Sciences"
of Eugene Wigner.

\item[${\cal Q}$] {\em (Yurij Holovatch)}: Is the Bohm theory the only known non-local hidden-variable theory?

\item[${\cal A}$] No, an alternative theory called stochastic mechanics was also pretty much
studied. In 1966, Nelson showed that the Schr\"odinger equation could be
derived from the hydrodynamical equation of a classical fluid made of particles
with Brownian trajectories (\paperBB{E. Nelson}{Phys. Rev.}{150}{1966}{1079}
	{Derivation of the Schroedinger Equation from Newtonian Mechanics}). 
He suggested that the interaction
of the particles with the vacuum for instance could induce fluctuations of their
position. Because of these fluctuations, the trajectories are continuous but
not differentiable. At a given time, two different velocities can be defined:
the advanced and the retarded velocities related to each other by reversing the
direction of time:      
	\begin{equation}
	\vac b(t)=\lim_{\Delta t\rightarrow 0^+} {\langle \vac r(t+\Delta t)
        -\vac r(t)\rangle\over\Delta t},\quad\quad
        \vac b_*(t)=\lim_{\Delta t\rightarrow 0^+} {\langle \vac r(t)
        -\vac r(t-\Delta t)\rangle\over\Delta t}
	\end{equation}
where $\langle\ldots\rangle$ denotes the average over the fluctuations.
At the macroscopic level, the density of particles satisfies a diffusion equation
with a diffusion constant depending on the Planck constant.
By reversing the direction of time, the sign of this diffusion constant changes.
Nelson combined these two equations to get a Navier-Stockes-like hydrodynamical
equation
	\begin{equation}
	{\hbar\over 2m}\Delta\vac u-\big(\vac u.\nabla\big)\vac u
        +\big(\vac v.\nabla\big)\vac v+{1\over m}\nabla V
        =-{\partial\vac v\over\partial t}
	\end{equation}
where
	\begin{equation}
	\vac v={1\over 2}\big(\vac b+\vac b_*\big),\quad
	\vac u={1\over 2}\big(\vac b-\vac b_*\big)
	\end{equation}
that can then easily be related to the Schr\"odinger equation with the wave
function
	\begin{equation}
	\psi=e^{R+iS},\quad
	\vac v={\hbar\over m}\nabla S,\quad
	\vac u={\hbar\over m}\nabla R
	\end{equation}
The stochastic mechanics still suffers from several problems. First of all, the
interpretation of the theory is not trivial even though some progresses have
been made by Fritsche and Haugk in 2003
(\paperBB{L. Fritsche et M. Haugk}{Ann. Phys. (Leipzig)}
	{12}{2003}{371}{A new look at the derivation of the Schrodinger
	equation from Newtonian mechanics}). Moreover, there is
still no relativistic generalization of the Nelson approach.
\par

\end{itemize}

\end{document}